\documentclass{PoS}
\usepackage{epsfig,multicol}

\newcommand{\beq}{\begin{eqnarray}}
\newcommand{\eeq}{\end{eqnarray}}
\newcommand{\diff}{\mbox{d}}

\PoS{PoS(LAT2005)053}

\title{Searching for diquarks in hadrons }

\ShortTitle{Searching for diquarks in hadrons }

\author{Constantia Alexandrou\\
  Department of Physics, University of Cyprus,
  CY-1678 Nicosia, Cyprus\\
  E-mail: \email{alexand@ucy.ac.cy}
}

\author{Philippe de Forcrand\\
  Institute for Theoretical Physics, ETH Z\"urich,
  CH-8093 Z\"urich, Switzerland\\
  CERN, Physics Department, TH Unit.
  CH-1211 Gen\`{e}ve 23, Switzerland\\
  E-mail: \email{forcrand@phys.ethz.ch}}

\author{\speaker{Biagio Lucini}\\
  Institute for Theoretical Physics, ETH Z\"urich,
  CH-8093 Z\"urich, Switzerland\\
  E-mail: \email{lucini@phys.ethz.ch}}

\abstract{
Since the early days of QCD, it has been argued that inside hadrons
quarks organise into substructures. Among those, popular pictures of various
phenomena (including exotica and colour superconductivity) give a prominent
role to diquarks, i.e. to colour antitriplet combinations of two quarks.
Using a gauge-invariant setup, which combines the diquark with a static quark,
we study spatial correlations of the two light quarks inside the diquark.
After
illustrating the setup, we discuss our first results for both the 
scalar (``good'') and the spin one (``bad'') diquark
channels. In a regime in which $m_{\pi} \simeq 830$ MeV, our data show
unambiguously that a scalar diquark forms with a size of $\simeq$ 0.9
fm. For the vector diquark we are able to put a lower bound of 4.1(7) fm on its
size. We also investigate the mass splitting between the good and the bad
diquark; our findings are compatible with phenomenologically inspired
predictions for this quantity.}

\FullConference{XXIIIrd International Symposium on Lattice Field Theory\\
                 25-30 July 2005\\
                 Trinity College, Dublin, Ireland}

\begin{document}

\section{Introduction}
\label{sect:introduction}
The static quark model has provided a rather accurate description of
hadronic states over the last four decades. This is a rather simple model,
since it assumes that quarks inside hadrons are uncorrelated.
Correlations among quarks could
however explain remaining features of the
spectrum that look incompatible with the simple-minded static quark model.
In fact a natural possibility is that two quarks combine into a colour
antitriplet representation. This correlation is energetically favourable,
since at large distances the resulting chromoelectric field is simply that of
an antiquark. Colour antitriplet combinations of two quarks are called
{\em diquarks}.

Diquarks have been discussed since the early days of QCD (see
Ref.~\cite{Anselmino:1992vg} for a review). Interest in this topic
has been recently revived by the increased attention devoted to the high
density phase of QCD~\cite{Alford:1998mk} and to the possible existence of
exotic particles~\cite{Jaffe:2003sg,Maiani:2004uc}, since in both cases
diquarks could play an important role. Diquark-inspired phenomenological models
usually start from the assumption that those bound states
exist. However this assumption needs to be justified from first
principles. As usual the lattice is an excellent framework for such {\em ab initio}
calculations. By using a gauge-invariant formalism that enables us to study
their structure, we  show unambiguously that diquark states do form and
present preliminary results for the mass splittings between
the scalar and vector  diquark channels. Our  calculation
is done in the quenched theory using Wilson fermions and quark
masses that correspond  to pion mass in the range  $830-1000$ MeV.

\section{Probing the structure of a diquark}
\label{sect:density}
Diquark operators have been discussed in~\cite{Jaffe:2004ph}. For convenience, the quantum numbers
and the colour and flavour representations of the simplest diquark operators
are reported in Table.~\ref{tab:1}. 
In this work we  focus on
the $\bar{q}_C \gamma_0 \gamma_5 q$ (the scalar diquark) and the
$\bar{q}_C \vec{\gamma} q$ (the vector diquark) combinations.
 Effective colour-spin hamiltonian arguments suggest that the spin zero
channel is enhanced with
respect to the spin one combinations~\cite{Jaffe:2004ph}. For this reason
the scalar and the vector diquarks are also referred to respectively as
the ``good'' and the ``bad'' diquarks. 
\TABLE[htb]{
  \begin{tabular}{|c|c|c|c|}
    \hline  
    $J^P$ & Colour & Flavour & Operator\\
    \hline
    \ & \ & \ & \ \\
    $0^{+}$ & $\bar{3}$ & $\bar{3}$ & $\bar{q}_C \gamma_5 q \ , \bar{q}_C \gamma_0 \gamma_5 q$\\
    \ & \ & \ & \ \\
    $1^{+}$ & $\bar{3}$ & 6 & $\bar{q}_C \vec{\gamma} q \ , \bar{q}_C \sigma^{0i} q$\\
    \hline
  \end{tabular}
  \caption{The structure of the simplest diquark operators ($\bar{q}_C$ is the
    conjugated spinor of $q$).}
  \label{tab:1}
}
We now want to investigate whether a diquark bound state forms.
The issue is complicated by the fact that diquarks
are coloured objects. In order to formulate the problem in a
gauge-invariant way, we combine the diquark with a static quark in a
colour singlet representation. Far from the source,
the influence of the static chromoelectric field becomes less important and the
physics of the diquark emerges.
The static-light baryon obtained by combining a
static colour source with a light diquark 
can be studied with the density-density correlator
method~\cite{Negele:2000uk}, which provides a gauge-invariant 
technique for studying quark distributions inside a hadron.
This method has been recently used in
Ref.~\cite{Alexandrou:2002nn,Choe:2003wx} 
for studying the structure of various hadronic
states.\\ 
\FIGURE[t]{
\epsfig{file=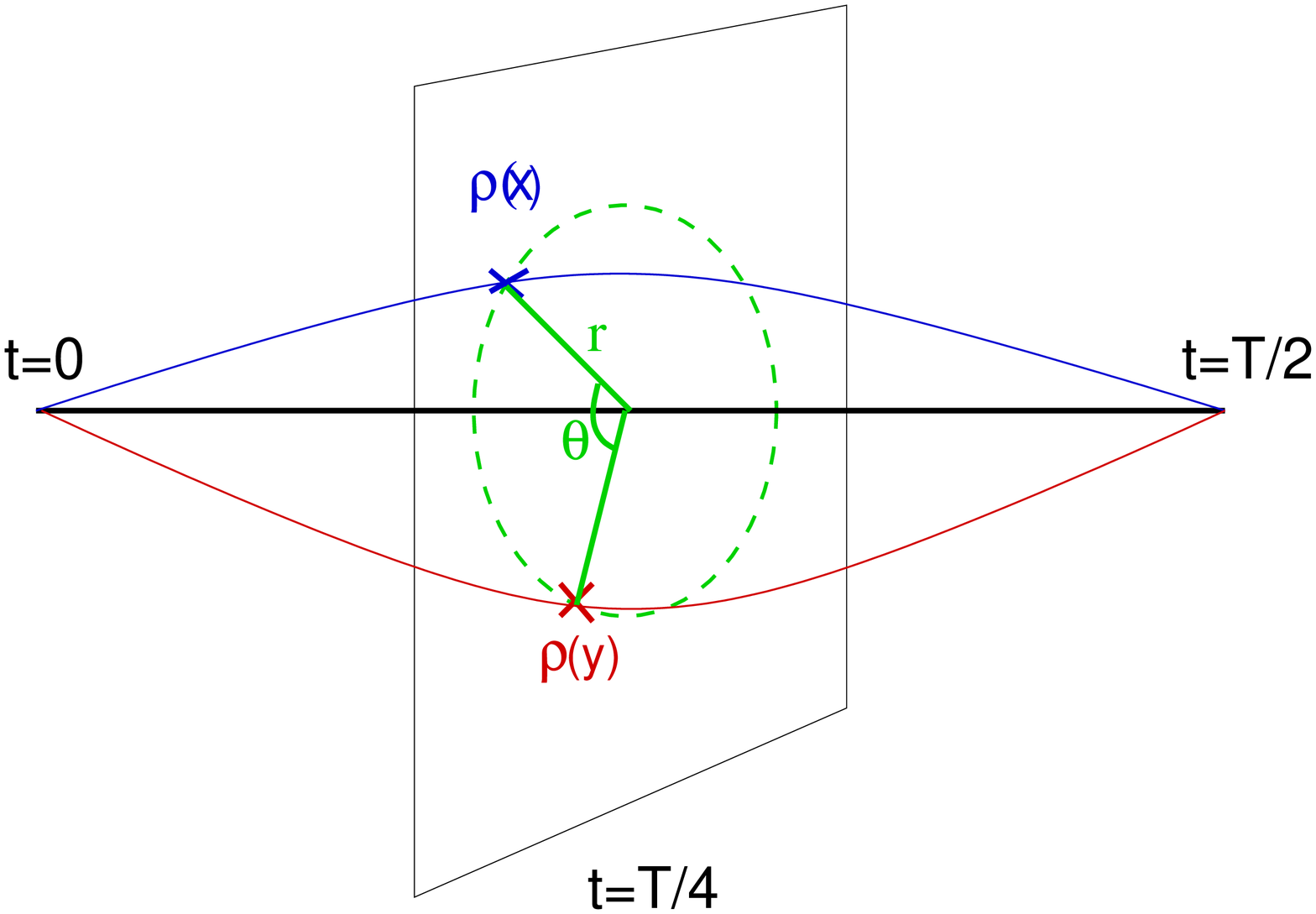,scale=0.4}
\caption{Setup for the computation of the density-density correlators.}
\label{fig:1}
}
Our setup is sketched in Fig.~\ref{fig:1}.
At Euclidean time $t=0$ we create a static-light baryon in which
the two light quarks are combined into a candidate diquark channel (either
scalar or vector) and we let this state propagate to time $t=T/2$ ($T$
is the extension of the lattice in the temporal direction), where it is
annihilated. Far from the source and the sink, at a fixed Euclidean time $t$ ($t=T/4$
in Fig.~1) where the hadron has relaxed to
its ground state, we study the density-density correlator of the two light
quarks
\beq
C(\theta, r) = \int \langle H| \rho_u(\vec{r}_u) \rho_d(\vec{r}_d)
| H \rangle \
\delta(|\vec{r_u}| - r) \delta(|\vec{r_d}| - r) \delta(\hat{r_u}\cdot\hat{r_d} - cos \theta) \diff\vec{r}_d \diff\vec{r}_u \ , 
\eeq
where the subscripts $u$ and $d$ refer respectively to the up and down
quarks, $\rho_q = \ :\bar{q} \gamma_0 q:$ and $\vec{r}_u$ and $\vec{r}_d$
are the lattice vector coordinates of the light quarks with
respect to the static source. To make the analysis of our results clearer, we
consider correlators for $|\vec{r}_u| = |\vec{r}_d| = r$: the two light
quarks are on the same spherical shell
at distance $r$ from the static source. Since the system is
spherically symmetric, $C$ only depends on $r$ and the angle $\theta =
\arccos(\hat{r}_u\cdot\hat{r}_d)$. If a diquark state is favored by  QCD
dynamics, one would expect that $C(\theta,r)$ is enhanced at small
values of $\theta$, while if the quarks
are uncorrelated the distribution is flat in $\cos\theta$. On a lattice
the angular distribution is distorted by discretisation
artifacts, especially for small values of $r$. To suppress those effects we have normalized our distributions to
that obtained from uncorrelated objects on the same lattice. \\
\FIGURE[t]{
~\\~\\
\epsfig{file=FIGS/diquark0_vs_5.eps,scale=0.4,angle=0}
\caption{$C(\theta,r)$ versus $\cos\theta$ at distance
  $r=5.1~a$ from the static source.}
\label{fig:2}
}
We have measured $C(\theta,r)$ on 200 quenched configurations at
$\beta = 6.0$ taken from the
NERSC archive\footnote{See http://qcd.nersc.gov.} using two degenerate
flavours of Wilson fermions at $\kappa = 0.153$. Those lattice parameters
correspond to $m_{\pi} \simeq 830$ MeV and to a lattice spacing 
$a \simeq 0.097$ fermi.
The results are obtained by binning in
$\cos\theta$ within shells of ``thickness'' $\delta r = 0.1 r$ (so that $|\vec{r}_u|/|\vec{r}_d| = 1 \pm 0.1$).
The magnitude of the correlator decays exponentially with the distance $r$ from the
static quark, with a decay length of 0.90(2) $a$
(corresponding to a mass of 2.25(5) GeV). Despite this, the signal never
dies into noise and we are able to study the density-density correlator
up to the maximal allowed distance $r = 8~a$ on our $16^3\times 32$ lattice.
In Fig.~\ref{fig:2} we report the angular distribution of $C(\theta,r)$
for a distance $r = 5.1~a$ from the static quark for  the scalar
diquark and the vector diquark  with the $z$ component of the spin
$J_z = 1$. An enhancement of the distribution at $\cos\theta = 0$ is
evident for the scalar diquark, while the distribution looks flatter for the
vector diquark.
A study of $C(r,\theta)$ at larger distances $r$ shows the same
features. This confirms the expectation that the color field generated by
the static source does not affect the physics of the diquark once $r$ is large enough,
and shows that this asymptotic regime can be reached in our calculations.
Hence the natural conclusion of this study is that a scalar diquark forms,
while a vector diquark state, if it exists, is much broader.
The density-density correlator technique allows also to give a
definition of the size of the diquark. At fixed $r$ we can look at the correlator 
as a function of the $u-d$ separation $R_{ud} = 2 r \sin (\theta/2)$. An exponential behaviour
$C(r,R_{ud}) \propto e^{-R_{ud}/R_0}$, as seen in ordinary hadrons~\cite{Alexandrou:2002nn},
gives a gauge-invariant definition of the hadron size $R_0$. For the scalar diquark, our data are
correctly described by an exponential ansatz over a wide range of $r$
(see Fig.~\ref{fig:3} for an example).
Results for $R_0$ as a function of $r$ are reported in Fig.~\ref{fig:4}.
At small separations $r$ the value of $R_0$ depends on the
radius $r$ of the shell.
The extracted values of $R_0$ converge for $r \ge 4~a$.
Once again, this confirms that
the regime in which the physics of the diquark system is not disturbed by the
field of the static source can be reached on our lattice. A safe estimate for
the scalar diquark size is $R_0 = 0.92(9)$ fm.
For the vector channel, the fitted size is noisier and seems to
increase with $r$. The value at the maximal possible distance $r = 8~a$ can be
used to put a lower bound of 4.1(7) fm on the size of this diquark.
\DOUBLEFIGURE[t]{FIGS/expfit_d6.15.eps,scale=0.28,angle=0}{FIGS/diquark0_size.eps,scale=0.28,angle=0}{The exponential decay of $C(r=6.15a,R_{ud})$ at fixed $r$ for the scalar diquark (the fit is performed from $R_{ud}=2~a$ to $8~a$). 
\label{fig:3}}{Fitted diquark size $R_0$ (in units of $a$) as a function
of the distance $r/a$ of the diquark from the static quark. \label{fig:4}}
\section{Diquark masses}
\label{sect:masses}
From colour-spin effective hamiltonian arguments one can estimate the
mass of the scalar diquark and the mass splitting 
$\Delta M = M_{\rm vector} - M_{\rm scalar}$ between the vector and the
scalar diquark, obtaining respectively  $M \simeq 320$ MeV and
$\Delta M \simeq 200$ MeV~\cite{Jaffe:2004ph}. 
Since we are using heavy quarks  we expect these numbers to be
different in our case. Again the
colour-spin effective hamiltonian gives us a handle on what we should expect:
if the mass of the degenerate constituent quarks increases, the mass $M$ of the
scalar diquark increases and the mass splitting $\Delta M$ goes down
as 1/$M^2$.
Within our static-light baryon setup the mass of the diquark can not be
accessed, since the mass 
of a hadron with a static quark is ultraviolet divergent. However,
the mass difference can be extracted by looking at the temporal behaviour of
the ratio of propagators of static-light baryons with the light quarks combined
into the channels we are interested in.
\TABLE[h]{
  \begin{tabular}{|c|c|c|c|}
    \hline  
    $\beta$ & $\kappa$ & $m_{\pi}$ (MeV) & $\Delta M$ (MeV)\\
      \hline
      6.0 & 0.1530 &  830 & 115(20)\\ 
      5.8 & 0.1575 & 830 & 100(15) \\ 
      5.8 & 0.1530 & 1000 & 67(7) \\   
      \hline
    \end{tabular}
    \caption{Mass splittings for the scalar and the vector diquarks at the
$\beta$ and $\kappa$ shown.}
\label{tab:2}
}
We  study these ratios on a
$16^3 \times 32$ lattice at a given pion mass of about 830 MeV
at $\beta=6.0$ taking $\kappa=0.153$  and $\beta = 5.8$ with $\kappa=0.1575$
to check scaling. At $\beta=5.8$ we also calculate $\Delta M$ at a larger pion
mass of about 1 GeV taking $\kappa=0.153$ to check the dependence of
$\Delta M$ on the quark mass. We use 200 NERSC configurations
for each value of $\beta$ and $\kappa$. To reduce the noise caused by the
static source we used hypercubic smearing~\cite{Hasenfratz:2001hp,DellaMorte:2005yc}.
Our results are reported in Table~\ref{tab:2}
where the errors given are only statistical. Changing the fitting range 
we obtain values for $\Delta M$ that are consistent within these large 
statistical errors. The following features emerge:
(a) at fixed $m_{\pi}$, the value of the mass splitting is 
independent of
$\beta$, and this suggests that our calculation is reasonably free of lattice
artifacts; (b) the mass splitting at $m_{\pi} = 830$ MeV is smaller than
the phenomenological estimate; (c) as $m_{\pi}$ is increased, the mass splitting
decreases. (b) and (c) are compatible with qualitative expectations from
colour-spin effective hamiltonian arguments above. Moreover, our results for the
mass splittings are also compatible with a similar calculation reported at
this conference~\cite{Poslat05054}, with estimates coming from an instanton
liquid model~\cite{Poslat05052} and with an earlier lattice
calculation in Landau gauge~\cite{Hess:1998sd} (see also~\cite{Babich:2005ay}).

\section{Conclusions}
\label{conclusions}
Density-density correlators in the background of a static source provide a
gauge-invariant method for the study of diquark structure.
Within this framework, we study the scalar and the vector diquark channels.
Our results show that a diquark state forms in the scalar channel, with a size
of  0.92(9) fm (for a pion mass around 830 MeV). For the vector diquark
we can provide a lower bound of 4.1(7) fm on its size. This is the first gauge
invariant calculation from first principles that  gives evidence for
the existence of diquarks as colour antitriplet bound states of two quarks.
The mass splitting between the scalar and vector diquarks is calculated;
our results are compatible with extrapolations of phenomenological estimates
towards large masses of the constituent quarks.
Currently we are simulating at a third value of $\beta$ to check scaling 
and be able to study heavier quarks. At the same time we are increasing the
statistics at $\beta=6.0$ and 5.8 by a factor of 10. By using three values of
quark masses we should be able to check the $1/M^2$ scaling of the mass
splitting between scalar and vector diquarks expected at heavy quark masses. 
Finally we are extending the study of the density-density
correlation functions and of the mass splitting to the other diquark
channels listed in Table~1.
\acknowledgments
We thank R. Jaffe and A. Polosa for useful discussions.
\bibliographystyle{JHEP}
\bibliography{diquarks}

\end{document}